\documentclass[a4paper]{spie} 

\usepackage[]{graphicx}

\title{The controlled indirect coupling between spatially-separated qubits in antiferromagnet-based NMR quantum registers}

\author{A.\ A.\ Kokin}
\date{}

\begin{document}

\maketitle

\thanks{Institute of Physics and Technology of RAS, 34, Nakhimovskii pr., 117218 Moscow, Russia}

\begin{abstract} \normalsize
It is considered the indirect inter-qubit coupling in 1D chain of atoms with 
nuclear spins 1/2, which plays role of qubits in the quantum register. This 
chain of the atoms is placed by regular way in easy-axis 3D 
antiferromagnetic thin plate substrate, which is cleaned from the other 
nuclear spin containing isotopes. It is shown that the range of indirect 
inter-spin coupling may run to a great number of lattice constants both near 
critical point of quantum phase transition in antiferromagnet of spin-flop 
type (control parameter is external magnetic field) and/or near homogeneous 
antiferromagnetic resonance (control parameter is microwave frequency). 
\end{abstract}


\section{Introduction}

In 1998 B.Kane proposed the scheme of large-scale NMR quantum register, 
based on nuclear spin-free substrate $^{28}$Si, into a near-surface layer of 
which atoms $^{31}$P, whose nuclear spins 1/2 play the role of qubits, are 
implanted in the form of a regular chain \cite{1}. One of the main quantum 
operations is a two-qubit operation, such as CNOT-operation, for which it is 
needed to switch on a coupling between considered qubits. It is assumed, 
that the indirect nuclear spin coupling for neighboring donors is controlled 
by special gates and two-qubit operations for far spatially separated qubits 
can be produced using SWAP operations between neighboring qubits. The 
separation between neighboring donors in this scheme must be $\sim $ 20 nm. 
A model of NMR quantum register based on two-leg ladder 1D antiferromagnet 
chain, where nuclear spins-qubits are placed in the magnetic field gradient 
and separated by a distance up to several tens of lattice constants, was 
proposed in [2]. The needed indirect inter-qubit coupling\textbf{} of 
Shul-Nakamura type in this model must be switched on by the external magnon 
packet excitations in the antiferromagnetic chain. Authors of \cite{2} assume 
some organic materials as the possible candidates for the NMR quantum 
register base.

In this paper, we consider a model of NMR quantum register, which is based 
on the easy-axis 3D antiferromagnet at low temperature. It is shown that the 
range of indirect coupling can range up to a great number of lattice 
parameters both near critical point of quantum phase transition in 
antiferromagnet of spin-flop type and/or near homogeneous antiferromagnetic 
resonance.

\section{Easy-exis antiferromagnets for nmr quantum registers}

\begin{figure}
   \begin{center}
   \begin{tabular}{c}
   \includegraphics{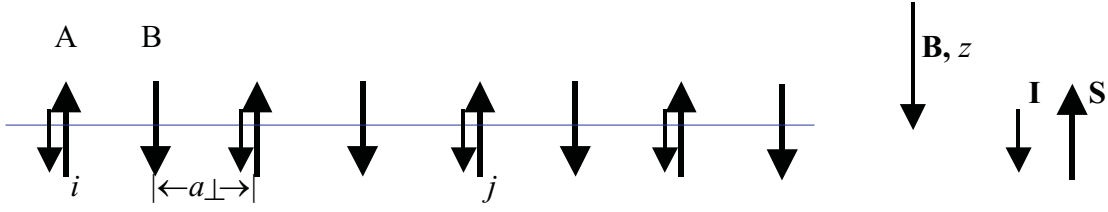}
   \end{tabular}
   \end{center}
   \caption[1]
%>>>> use \label inside caption to get Fig. number with \ref{}
   { \label{fig:1} The scheme of antiferromagnet based NMR quantum register. The nuclear spin is contained only in atoms A.
   }
   \end{figure}

We propose here that 1D chain of atoms with nuclear spins is placed in 
easy-axis 3D antiferromagnetic thin substrate-plate ($d$ is the thickness), 
which is cleaned from the other nuclear spin containing isotopes (Fig.~\ref{fig:1})

%-------------
%-------------

Consider next a symmetric easy-axis antiferromagnetic structure 
(rhomboedric, trigonal or hexagonal crystal symmetry). The qubits with 
numbers $i$ and $j$ are separated by distance $x_{ij} \equiv a(i-j)$
($a$ is lattice parameter) along the plate. As examples, the 
following well-known easy-axis natural crystals FeCO$_{3}$ (siderite), 
á-Fe$_{2}$O$_{3} $(hematite), CeC$_{2}$, FeGe, FeGe$_{2}$ may be probably 
proposed. The isotopes $^{13}$C, $^{57}$Fe are here as nuclear spin 
containing atoms when they substitute for the definite host spin-free atoms 
in plate. The indirect inter-qubit coupling is due to hyperfine interaction 
of nuclear and electron spins in substituted atom, coupling of that electron 
spin is mainly through exchange interactions with electron spin of the 
neighboring host atom and spin-wave (magnon) propagation that is caused by 
the exchange interaction between electron spins of host atoms. The external 
homogenous field $\mathbf{B}$ (z-axis) is directed normally to the surface of plate 
and in parallel with the easy-axis.

Let us use for an antiferromagnet model the system of two magnetic 
sublattices A and B with $L/2$ sites in each sublattice ($L$ is even full site number).
The sites will be numbered by numbers $i$ and $j$ accordingly.
The electron spin Hamiltonian for our model for 
3D easy-axis antiferromagnets with interaction only between neighboring 
atoms may be represented as (in frequency units)

\begin{eqnarray}
\nonumber
&
H_{S} = \gamma _{S} B\;\left( {\;\sum\limits_{i}^{L/2} {S_{Az}(i)} + \sum\limits_{j}^{L/2} {S_{Bz}}(j)} \right) +
&
\\
& + 2\gamma _{S} /Z\;\;\sum\limits_{i}^{L/2} {\;\sum\limits_{\delta}^{Z}
{\{B_{E} \mathbf{S}_{A} (i) \;\mathbf{S}_{B}(i+ \delta) \; + \;B_{A} \;S_{Az}(i)\;S_{Bz}(i + \delta)\}} },
&
\label{eq1}
\end{eqnarray}

\noindent
where $\mathbf{S}_{A} $ and $\mathbf{S}_{B}$ are electron spin operators,
$Z$ is the number of neighboring atoms, $B_{E} > B_{A} \ge 0$ are exchange and anisotropy fields for easy-axis 
antiferromagnet, $\gamma _{S} = 176.08$ GHz/T is gyromagnetic ratio for 
electron spin.

Under low temperature conditions, the electron spins are essentially in 
ground state and deviations of their states from ground state are small and 
may be considered in spin-wave approximation.

Spin-wave (magnon) Hamiltonian after diagonalization, as it is known \cite{3}, 
takes the form

\begin{eqnarray}
\label{eq2}
H_{S} \approx \textrm{Const} + \sum\limits_{\mathbf{k}}
{\left( {\omega_{+}(\mathbf{k},B)\;N_{+}(\mathbf{k}) + \omega_{-}(\mathbf{k},B)\;N_{-}(\mathbf{k}}
\right),
}
\end{eqnarray}

\noindent
where $\mathbf{k}$ is spin-wave vector, the values of magnon number
$N_{\pm}(\mathbf{k})$ are 0 or 1. The long-wave magnon frequencies
($k_{\bot} a_{\bot},\;k_{z} a_{z} \ll 1$), where $k_{\bot}$ is radial component, are

\begin{eqnarray}
\label{eq3}
\omega_{\pm}(k,B) \approx \gamma _{S}
\left( {\sqrt{\xi ^{2}B_{E}^{2}
\left( {a_{\bot}^{2} k_{\bot}^{2} + a_{z}^{2} k_{z}^{2}}\right)
+ 2B_{A} B_{E} + B_{A}^{2}}  \pm B} \right),
\end{eqnarray}

\noindent
where $\xi^{2}\sim 1$, $a_{\bot}$ and $a_{z}$ are lattice parameters of
the order of lattice constants.

We see that magnon spectrum of easy-axis antiferromagnetic has energy gap

\begin{eqnarray}
\label{eq4}
\omega_{-}(0,B) > 0
\end{eqnarray}

\noindent
for fields $B < B_{C} = \sqrt {2B_{A} B_{E} + B_{A}^{2}}$, where value of
$B_{C}$ is in the neighborhood of 10 tesla

Under low temperature condition ($k_{B}$ is Boltzmann constant)

\begin{eqnarray}
\label{eq5}
T \ll \hbar \omega_{-} \left( {0,B} \right)/k_{B} = T_{C} \left( {1 - B/B_{C}}  \right),
\end{eqnarray}

\noindent
where $T_{C} = \hbar \gamma _{S} B_{C} /k_{B} $ is less than 10 K, the 
easy-axis antiferromagnet conserves homogeneous ground state. This condition 
provides also a very long transverse nuclear spin relaxation time \cite{4} and 
accordingly the one qubit decoherence time.

If the field $B$(control parameter) goes through the critical field $B_{C} 
$, the stability of grand state breaks down (Goldstone instability) and the 
quantum phase transition in the homogeneous $(\mathbf{k}= 0)$ 
spin-flop phase occurs.

Let us move now to condition of antiferromagnetic resonance. In the presence 
of the rotating transverse microwave field with frequency $\omega$ (control 
parameter), the $B$ is replaced by

\begin{eqnarray}
\label{eq6}
B \to B_{\mathrm{eff}}(\omega) = (B + \omega /\gamma_{S}).
\end{eqnarray}

The lower magnon frequency in rotating frame takes now the form

\begin{eqnarray}
\label{eq7}
\omega_{-} (\mathbf{k},B,\omega) \approx \gamma _{S} \;\left( 
{\sqrt{\xi^{2}B_{E}^{2} \left( {a_{\bot}^{2} k_{\bot}^{2} + a_{z}^{2}
k_{z}^{2}}  \right) + B_{C}^{2}} - B_{\mathrm{eff}}(\omega)} \right).
\end{eqnarray}

The homogeneous ($\mathbf{k}=0$) antiferromagnetic resonance corresponds to phase
transition due to instability in rotating frame:

\begin{eqnarray}
\label{eq8}
\omega_{-}(0,B,\omega) = 0, \mathrm{or} \omega = \gamma_{S}(B_{C}-B).
\end{eqnarray}

Let us assume that antiferromagnetic resonance width obeys the conditions
$\Delta \omega < \gamma _{S} B_{C} \ll \omega_{0} =
\gamma _{S} \xi ^{2}B_{E}^{2} /B_{C}$
and thickness $d$ of
infinitely large plate is $d < \pi a_{z} \left( {\omega_{0} /\Delta
\omega_{0}}  \right)^{1/2}$, that is, in neighbourhood of nanometers. Then
the 2D (along the plate surface, $k_{\bot} \ne 0,\;k_{z} = 0$) spin waves
are defined as homogeneous electron spin precession normally to the plate
surface or as zero mode of spin-wave (magnon) resonance.

\begin{eqnarray}
\label{eq9}
\omega_{-} \left( {k_{\bot}, \;k_{z}=0, \;B, \omega} \right) = 0,
\end{eqnarray}

\noindent
and we have the value for microwave frequency

\begin{eqnarray}
\label{eq10}
\omega \approx \gamma _{S} \;\left( {\sqrt{\xi ^{2}B_{E}^{2} a_{\bot}^{2} k_{\bot}^{2} + B_{C}^{2}}  - B} \right).
\end{eqnarray}

For homogeneous microwave field and for large plate wave vector $k_{\bot}
$ take the continuous values $0 \le k_{\bot} < \infty$, therewith value
$k_{\bot}=0$ corresponds to homogeneous antiferromagnetic
resonance in 2D structure (thin plate).

\section{Controlled indirect inter-qubit coupling in easy-axis antiferromagnet}

The indirect inter-spin coupling in chain is due to hyperfine interaction of
nuclear and electron spins in substituted atoms A in chain $A( \mathbf{I}_{A}
\mathbf{S}_{A})$, where $A$ is isotropic hyperfine coupling constant of the
order of 100 MHz. Electron spins of the neighboring host atom A and B are 
coupled mainly through the exchange interactions, which is responsible for 
2D spin waves (magnons) formation in particular with the frequency
$\omega_{-}(\mathbf{k},B)$. The nuclear spin of an atom A exits the
magnon, which is absorbed by nuclear spin of another atom A. (or B). In this 
way, the coupling between nuclear spins of separated atoms is produced. Note 
that no external spin wave excitation is involved here.

We restrict our consideration to the indirect inter-nuclear interaction at 
low temperature. The corresponding Shul-Nakamura Hamiltonian for two 
spins in one sublattice A $i$ and $j$ are located 
along $x$-axis is nonzero only for transverse component \cite{4}:

\begin{eqnarray}
\label{eq11}
H_{I} = - I_{\bot}(i-j)\;\left( {I_{Ax}(i)I_{Ax}(j) + I_{Ay}(i)I_{Ay}(j)} \right)
- \gamma _{S} \;B\;\left( {I_{Az}(i)
+ I_{Az}(i)} \right),
\end{eqnarray}

\noindent
where value of indirect nuclear inter-spin coupling have form

\begin{eqnarray}
\label{eq12}
I_{ \bot}  \left( {i - j} \right) = \;\;A\;^{2}/2N \cdot
\sum\limits_{k} {\exp \left( {k_{\bot} a_{\bot} (i - j)
\cos \varphi } \right)/\omega_{-}  \left( {k_{\bot},k_{z}=0,\;B,\;\omega} \right)},
\end{eqnarray}

\noindent
where $N$ is the full number of atoms in the plate. The function (\ref{eq11})
represents essentially a correlation function for transverse components of
electron spin operators.

We assume next that for large value of distance $|i - j|$ the dominant 
contribution gives small values of radial spin vectors $k_{\bot}  a_{ \bot
} < \left( {B_{C} /\xi B_{E}}  \right) \ll 1$. Then we are coming from
summation to integration over polar angle $\varphi$ and $k_{\bot}$ and obtain

\begin{eqnarray}
\nonumber
&
I_{ \bot}  \left( {i - j} \right)
\approx 
 \;\;A\;^{2}a_{z} a_{ \bot} ^{2}
/\left( {\gamma _{S} 4\pi \;d} \right) \cdot \int\limits_{0}^{\infty}{J_{0} \left( {k_{\bot} a_{\bot}(i - j)}\right)
/\left( {B_{C} - B_{\mathrm{eff}} (\omega)\; + \xi
^{2}B_{E}^{2} a_{\bot}^{2} k_{\bot}^{2} /2B_{C}} \right)\;k_{\bot} dk_{\bot} } =
&
\\
&= \;\;A\;^{2}a_{z} /\left( {\gamma _{S} 4\pi d} \right) \cdot
\int\limits_{0}^{\infty}  {J_{0} \left( {\zeta \left( {i - j} \right)}
\right)\;/\left( {B_{C} - B_{\mathrm{eff}} (\omega) + \xi
^{2}B_{E}^{2} \zeta ^{2}/2B_{C}}  \right)\;\zeta \;d\zeta}  ,
&
\label{eq13}
\end{eqnarray}

\noindent
where $J_{0} \left( {x} \right)$ is the Bessel function.

We now obtain

\begin{eqnarray}
\label{eq14}
I_{\bot}(i - j) \approx A^{2}a_{z}\; B_{C}/
\left({\gamma _{S} 2\pi d\xi ^{2}B_{E}^{2}}\right) \cdot K_{0}
\left({|i-j|/\rho_{B}(\omega)}\right),
\end{eqnarray}

\noindent
where $K_{0} \left( {x} \right)$ is McDonald function and for
$\;\;B_{\mathrm{eff}} (\omega) \le B_{C}$

\begin{eqnarray}
\label{eq15}
\rho _{B} (\omega) = \rho _{0} /\;\left( {1 - B_{\mathrm{eff}} \left( 
{\omega}  \right)/B_{C}}  \right)^{1/2},
\end{eqnarray}

\noindent
denotes the controlled range of indirect nuclear spin coupling (correlation 
length), $\rho _{0} = \xi B_{E} /\sqrt {2} B_{C} $ may be of the order of 
ten.

Let us consider two simple cases:

\noindent
a) Away from the point of phase transition and antiferromagnetic resonance 
$B_{\mathrm{eff}} (\omega)/B_{C} \ll 1,\;\;\rho _{B} \left( {\omega}  
\right) = \rho _{0} $ we obtain

\begin{eqnarray}
\label{eq16}
\left( {2\pi d\;\xi ^{2}\;\gamma _{S} B_{E}^{2} /A\;^{2}a_{z} B_{C}}
\right) \cdot \;I_{ \bot}  \left( {i - j} \right) \approx \;\left( {\pi \rho
_{0} /2|i - j|} \right)^{1/2}exp\left( {-|i-j|/\rho_{0}} \right) \ll 1
\end{eqnarray}

The indirect inter-spin coupling is relatively weak for any one
$|i-j| \ge 10 \rho_{0}$.

\noindent
b) Both near electron quantum phase transition
$(\omega = 0)$ and near antiferromagnetic resonance
$(\omega \ne 0$ for 
$\left( {1 - B_{\mathrm{eff}} (\omega)/B_{C}}  \right) \ll 1$ 
we have

\begin{eqnarray}
\label{eq17}
\rho _{0} \ll \rho _{B} (\omega);\;\;|i - j|/\rho _{B} 
(\omega) \ll 1:
\end{eqnarray}

\noindent
and

\begin{eqnarray}
\label{eq18}
\left( {2\pi d\;\xi ^{2}\;\gamma _{S} B_{E}^{2} \;/A\;^{2}a_{z} B_{C}}
\right) \cdot I_{\bot}(i-j) \approx \log \left(
{2\rho _{B} (\omega)/|i - j|} \right) \gg 1.
\end{eqnarray}

We see that increase of correlation length is attended with a decrease of
energy gap. However, the temperature therewith must correspond to low values
$T \ll T_{C} \;\left( {1 - B_{\mathrm{eff}}(\omega)/B_{C}} \right)$.
As a result, the controlled indirect coupling may be large for
distances $|i - j|$ as much as great number of lattice parameters.

\section{Conclusion}

We have shown that there is possibility to control the range inter-qubit 
coupling in easy-axis antiferromagnet-based 1D quantum registers through the 
correlation length variation near the quantum phase transition and/or near 
the antiferromagnetic resonance conditions. Our model of 1D nuclear quantum 
register corresponds to ferromagnetic transverse XX spin chain, but with not 
only pair interaction. It was shown that the nuclear two-spin correlation 
length may be raised to a great number of lattices constant.

We believe that the two-qubit operations, coherence state transfer and 
quantum states entanglement of far-separated qubits in the quantum register 
can be performed through similar long-range indirect coupling. It is not 
unlikely that entanglement may appear also for quantum states of opposite 
ends of spin chain giving the channel for entanglement transport across the 
chain.

We can consider also an ensemble variant of quantum register, which is made 
possible by the use of many moved apart in parallel working 1D spin chains.

The tuning of resonance frequencies of involved in quantum operation qubits 
may be performed through the imposition of local electrical gates. We do not 
discuss here special features of two-qubit decoherence for the far-separated 
spins-qubits states. Certain of related problems were discussed earlier in 
our book \cite{5}.

Note, that the external control of spin-wave excitations and propagation can 
play the role of the local measurements on the assisting electron spins, as 
this is required for localizable entanglement formation in 1D quantum 
registers with near neighbor interactions [6].

\end{document}